\documentstyle[amsmath,a4,11pt]{article}

\date{\empty}

\begin{document}

\title{\bf Resonant amplification of magnetic
seed fields by gravitational waves in the early universe}

\author{{Christos G.
Tsagas\thanks{email:tsagas@astro.auth.gr}}\\ {\small Section of
Astrophysics, Astronomy and Mechanics, Department of Physics}\\
{\small Aristotle University of Thessaloniki, Thessaloniki 54124,
Greece}\\ {\small DAMTP, Centre for Mathematical Sciences,
University of Cambridge}\\ {\small Wilberforce Road, Cambridge CB3
0WA, UK}}

\maketitle

\begin{abstract}
Inflation is known to produce both gravitational waves and seed
magnetic fields on scales well beyond the size of the horizon. The
general relativistic study of the interaction between these two
sources after the end of inflation, showed a significant
amplification of the initial magnetic seed which brought the
latter within the currently accepted dynamo limits. In the present
article we revisit this gravito-magnetic interaction and argue
that the observed strong growth of the field is the result of
resonance. More specifically, we show that the maximum magnetic
boost always occurs when the wavelength of the inducing
gravitational radiation and the scale of the original seed field
coincide. We also look closer at the physics of the proposed
Maxwell-Weyl coupling, consider the implications of finite
electrical conductivity for the efficiency of the amplification
mechanism and clarify further the mathematics of the analysis.\\\\
PACS number(s): 98.80.Hw, 04.30.-w, 98.80.Cq
\end{abstract}

\section{Introduction}
Observations have repeatedly verified the widespread presence of
magnetic fields in the universe~\cite{K}. Large-scale fields have
been found in galaxies, galaxy clusters, superclusters and also in
high-redshift radio galaxies. The typical magnetic strengths vary
between a few and several $\mu$G, while the associated coherence
lengths are comparable to those of the virialised hosts. Despite
their ubiquitous presence, however, the origin of these fields is
still a matter of open debate~\cite{E}. Over the years many
scenarios of cosmic magnetogenesis have appeared in the
literature. These range from eddies, density fluctuations and
reionisation effects in the post-recombination plasma to
cosmological phase-transitions, inflationary and
superstring/M-theory inspired scenarios~\cite{H1}. Historically,
the study of magnetic generation has been motivated by the need to
explain the origin of the large-scale galactic fields. The
structure of these fields, particularly those seen in spiral
galaxies, supports the galactic dynamo idea~\cite{P}. Although the
efficiency of the mechanism has been criticised, it is generally
believed that galactic dynamos can substantially amplify
preexisting weak magnetic seeds. The origin of the seed fields,
however, is still elusive. When the dynamo amplification is
efficient, the initial field can be as low as $\sim10^{-23}$~G at
the time of completed galaxy formation~\cite{Ku}. This limit is
relaxed to $\sim10^{-30}$~G in universes dominated by dark
energy~\cite{DLT}. In the absence of dynamo, however, magnetic
seeds of the order of $10^{-12}$~G, or even $10^{-8}$~G, are
required. The scale of the initial field is also an issue, since
galactic dynamos require a minimum coherence length of
$\sim100$~pc to guarantee the stability of the process~\cite{KA}.
In summary, the lowest current theoretical requirement for the
dynamo to work is a magnetic seed close to $10^{-30}$~G on a
collapsed scale of $\sim100$~pc. This corresponds to a field of
approximately $10^{-34}$~G with a comoving length of roughly
10~kpc.

The possible cosmological origin of the initial magnetic seeds is
an appealing suggestion because it can explain both the fields
seen in nearby galaxies and those detected in galaxy clusters and
high-redshift condensations. Causality, however, means that the
coherence length of any field generated between inflation and
(roughly) recombination is unacceptably small. A mechanism known
as `inverse cascading' can solve this problem~\cite{C}, but it
requires large amounts of helicity to operate successfully.
Inflation has long been suggested as a solution to the causality
problem because it naturally achieves correlations on superhorizon
scales. The problem with inflation is that the residual magnetic
field is far too weak to sustain the galactic dynamo. The reason
is the `adiabatic', $a^{-2}$ depletion rate of the field ($a$ is
the cosmological scale factor) during the de Sitter phase. One can
get around this obstacle by slowing down the decay of the
primordial seed. The effect is known as `superadiabatic
amplification' and it is usually achieved by breaking away from
classical electromagnetic theory~\cite{TW}. There are more than
one ways of doing that, which explains the variety of the proposed
mechanisms in the literature~\cite{GFC,DDPT}.

It should be noted that when the FRW model has open spatial
curvature, the coupling between the field and the background
geometry can slow down the magnetic decay without violating
classical electromagnetism~\cite{TK}. This occurs during the
poorly conductive phase of de Sitter inflation and affects lengths
close to the curvature scale and beyond. As a result, magnetic
fields spanning those lengths decay as $a^{-1}$ (or slower)
instead of $a^{-2}$. Even if the universe is only marginally open
today, the mechanism can produce large-scale fields with
astrophysically interesting strengths. For example, assuming
$1-\Omega\sim10^{-2}$ at present, GUT-scale inflation and a
reheating temperature of $\sim10^{9}$~GeV, one obtains a residual
field of the order of $10^{-35}$~G on a scale $\sim10^4$~Mpc
today~\cite{TK}. Moreover, the aforementioned final strength can
increase by lowering the reheating temperature. Therefore,
breaking away from Maxwell's theory is not a necessary requirement
for the superadiabatic amplification of cosmological magnetic
fields in perturbed FRW universes.

A common feature amongst almost all inflationary models is the
production of gravitational radiation over a wide range of
wavelengths. The interaction of these waves with large-scale
magnetic fields soon after the end of inflation was originally
considered in~\cite{TDM}. That study showed that the
gravitationally induced shear can amplify the initial magnetic
seed and the boost was found to be proportional to the square of
the field's initial scale. The latter immediately suggested that
large-scale primordial magnetic fields could be substantially
amplified by Weyl curvature distortions alone. Seed fields
spanning a current scale of $\sim10$~kpc, like those produced
in~\cite{DDPT} for example, were boosted by up to 14 orders of
magnitude. In the present paper we revisit the aforementioned
gravito-magnetic interaction and discuss the mathematics and the
physics of the mechanism in further detail. We argue that the
achieved strong magnetic growth results from the resonant coupling
of the two interacting sources. More specifically, we show that
the maximum amplification always occurs when the original seed
field interacts with gravitational waves of the same scale. The
maximum boost is determined at the onset of the gravito-magnetic
interaction, which for our purposes coincides with the end of
inflation. Once the parameters of the adopted inflationary model
are fixed, the resonant growth factor is proportional to the
initial magnetic scale, relative to the horizon size at the time.
Also the whole process is shown to operate in cosmological
environments of low electrical conductivity. All these make the
proposed amplification mechanism a highly efficient `geometric
dynamo' during the early stages of reheating. In this respect, the
Maxwell-Weyl resonance discussed here resembles the magnetic
amplification via parametric resonance proposed in~\cite{M,BPTV}.
Finally, given that the universe has been a good conductor for
most of its lifetime, we examine the role of finite conductivity
and establish the threshold at which the electrical resistivity of
the medium becomes unimportant. In agreement with the numerical
results of~\cite{BPTV}, our analytical approach suggests that the
gravito-magnetic resonance is suppressed is highly conductive
cosmological environments.

In sections 2 and 3 we provide a description of the model, of the
Maxwell-Weyl interaction and of the resulting magnetic
amplification. We follow the presentation of~\cite{TDM}, where the
reader is referred to for details, and provide additional
mathematical clarifications and physical insight. The
gravito-magnetic resonance is shown in section 4 and an expression
for the resonant growth factor is given. Section 5 applies the
proposed amplification mechanism to several inflation produced
magnetic seeds, while the role of finite electrical conductivity
is discussed in section 6.

\section{Gravito-magnetic interaction}
\subsection{Background equations}
Consider a spatially flat FRW universe containing a barotropic
perfect fluid of density $\rho$ and isotropic pressure
$p=p(\rho)$. Following~\cite{TDM}, allow for the presense of a
weak magnetic field $B_a$ with $B^2\ll\rho$. At this limit, the
field has negligible contribution to the background dynamics,
which is described by
\begin{equation}
\kappa\rho- {\textstyle{1\over3}}\Theta^2=0\,, \hspace{15mm}
\dot{\Theta}+ {\textstyle{1\over3}}\Theta^2+
{\textstyle{1\over2}}\kappa\rho(1+3w)=0\,,  \label{eq:Fried-Ray}
\end{equation}
\begin{equation}
\dot{\rho}+ (1+w)\Theta\rho=0\,, \hspace{15mm} {\rm and}
\hspace{15mm} \dot{B}_a+ {\textstyle{2\over3}}\Theta B_a=0\,.
\label{eq:edc-M1}
\end{equation}
In the above $\Theta=3\dot{a}/a=3H$ is the expansion scalar ($H$
is the Hubble parameter) and $w=p/\rho$ is the barotropic index of
the fluid~\cite{TDM}. Also, throughout this paper we use natural
units with $c=1=\hbar$ and $G^{-1/2}=m_{Pl}\simeq10^{19}$~GeV.

We perturb the aforementioned background by allowing for the
propagation of weak gravitational waves, which are covariantly
monitored via the electric ($E_{ab}$) and the magnetic ($H_{ab}$)
Weyl components~\cite{El}. In the magnetic presence, one isolates
the linear pure-tensor perturbations by imposing the conditions
${\rm D}_aB^2=0=\varepsilon_{abc}B^b{\rm
curl}B^c$~\cite{MTU}.\footnote{The gradient ${\rm
D}_a=h_a{}^b\nabla_b$, with $h_{ab}=g_{ab}+u_au_b$, is the
covariant derivative operator orthogonal to the observer
4-velocity $u_a$. Also, $\text{curl}B_a=\varepsilon_{abc}{\rm
D}^bB^c$, where $\varepsilon_{abc}$ is the spatial permutation
tensor (i.e.~$\varepsilon_{abc}u^a=0$). For more details and an
extensive covariant discussion of cosmological magnetic and
electromagnetic fields the reader is referred to~\cite{TB,T}.}
These, together with the standard constraints of the perfect-fluid
models (e.g.~see~\cite{Ch}), guarantee that all traceless
second-rank tensors are also transverse.

\subsection{Linear equations}
Adopting the aforementioned weakly magnetised FRW background, we
find that the linear magnetic evolution in the presence of
gravity-wave perturbations is governed by the system\footnote{The
most straightforward derivation of Eq.~(\ref{eq:ddotB1}) is by
linearising the nonlinear magnetic wave equation given in~\cite{T}
(see Eq.~(40) there). On the other hand, one can obtain
Eq.~(\ref{eq:ddotsig1}) directly from expression (3.11)
in~\cite{Ch}.}
\begin{equation}
\ddot{B}_a+ {\textstyle{5\over3}}\Theta\dot{B}_a+
{\textstyle{1\over3}}(1-w)\Theta^2B_a- {\rm D}^2B_a=
2\left(\dot{\sigma}_{ab}
+{\textstyle{2\over3}}\Theta\sigma_{ab}\right)\tilde{B}^b+ {\rm
curl}{\cal J}_a\,,  \label{eq:ddotB1}
\end{equation}
\begin{equation}
\ddot{\sigma}_{ab}+ {\textstyle{5\over3}}\Theta\dot{\sigma}_{ab}+
{\textstyle{1\over6}}(1-3w)\Theta^2\sigma_{ab}- {\rm
D}^2\sigma_{ab}= 0\,,  \label{eq:ddotsig1}
\end{equation}
where ${\cal J}_a$ is the 3-current and $\sigma_{ab}$ is the
gravitationally induced shear~\cite{TDM}. Note that $\tilde{B}_a$
is the original magnetic field, and $B_a$ is the new one induced
by the coupling between $\tilde{B}_a$ and gravity wave
distortions. The induced field vanishes in the background, which
frees our study from potential gauge-related ambiguities. In
addition, to first order, only $\tilde{B}_a$ contributes to the
right-hand side of (\ref{eq:ddotB1}).

At this stage we will ignore the current term in
Eq.~(\ref{eq:ddotB1}). This confines our analysis to a medium of
zero electrical conductivity or to an electrically neutral
one.\footnote{Mathematically speaking, the gravito-magnetic
interaction is independent of the electrical resistivity if ${\rm
curl}{\cal J}_a=0$~\cite{TDM}. Physically, however, the assumption
of a curl-free current field appears rather arbitrary and we will
not pursue it here.} In section 6, however, we will show that our
results also hold in cosmological environments of finite but
relatively low electrical conductivity. We have also ignored the
magnetic backreaction in (\ref{eq:ddotsig1}) because it does not
affect the dominant linear mode of the gravitationally induced
shear~\cite{MTU}. Finally, we note that gravity-wave perturbations
are the driving force behind the magnetic adulation described by
Eq.~(\ref{eq:ddotB1}). In particular, one can explicitly show that
the Weyl field alone triggers fluctuations in an otherwise
homogeneous magnetic field distribution (see~\cite{TB}).

Expression (\ref{eq:edc-M1}b) means that
$\tilde{B}_a=\tilde{B}_a^0(a_0/a)^2$, with
$\dot{\tilde{B}}_a^0=0$. By splitting the zero-order field as
$\tilde{B}_a= \tilde{B}_{({\rm n})}\tilde{Q}_a^{({\rm n})}$, we
assign the finite physical scale $\lambda_{\tilde{B}}=a/{\rm n}$
to $\tilde{B}_a$~\cite{TDM,MDB}. This, however, does not mean that
the background field is treated as a propagating wave of any sort.
Then, for $\sigma_{ab}=\sigma_{({\rm k})}Q^{({\rm k})}_{ab}$ and
$B_a=B_{(\ell)}Q_a^{(\ell)}$, where $Q^{({\rm k})}_{ab}$ and
$Q_a^{(\ell)}=Q_{ab}^{({\rm k})}\tilde{Q}_{({\rm n})}^b$ are
tensor and vector harmonics respectively, we have~\cite{TDM}
\begin{equation}
\ddot{B}_{({\ell})}+
{\textstyle{5\over3}}\Theta\dot{B}_{({\ell})}+
\left[{\textstyle{1\over3}}(1-w)\Theta^2
+\frac{\ell^2}{a^2}\right]B_{({\ell})}= 2\left(\dot{\sigma}_{({\rm
k})}+ {\textstyle{2\over3}}\Theta\sigma_{({\rm
k})}\right)\tilde{B}_0^{({\rm n})}\left(\frac{a_0}{a}\right)^2\,.
\label{eq:ddotB2}
\end{equation}
\begin{equation}
\ddot{\sigma}_{({\rm k})}+
{\textstyle{5\over3}}\Theta\dot{\sigma}_{({\rm k})}+
\left[{\textstyle{1\over6}}(1-3w)\Theta^2+\frac{{\rm
k}^2}{a^2}\right]{\sigma}_{({\rm k})}=0\,.  \label{eq:ddotsig2}
\end{equation}
Here, the zero suffix indicates the onset of the gravito-magnetic
interaction. Also, $\ell=({\rm k}^2+{\rm n}^2+2{\rm k}{\rm
n}\cos\varphi)^{1/2}$  is the comoving wavenumber of the induced
magnetic field and $\varphi\in[0,\pi/2)$ is the angle between the
two interacting sources. Finally, setting ${\cal
B}_{({\ell})}=\kappa^{1/2}B_{({\ell})}/\Theta$,  $\Sigma_{({\rm
k})}=\sigma_{({\rm k})}/\Theta$, using conformal time ($\eta$,
with $\dot{\eta}=a^{-1}$) and primes to indicate differentiation
with respect to $\eta$, the above recast as~\cite{TDM}
\begin{eqnarray}
{\cal B}''_{({\ell})}+ (1-3w)\left(\frac{a'}{a}\right){\cal
B}'_{({\ell})}-
\left[{\textstyle{3\over2}}(1-3w)w\left(\frac{a'}{a}\right)^2
-\ell^2\right]{\cal B}_{({\ell})}&=&\nonumber\\
2\kappa^{1/2}a\left[\Sigma'_{({\rm k})}
+{\textstyle{1\over2}}(1-3w)\left(\frac{a'}{a}\right)
\Sigma_{({\rm k})}\right]\tilde{B}_0^{(n)}
\left(\frac{a_0}{a}\right)^2\,,  \label{eq:cB''}
\end{eqnarray}
and
\begin{equation}
\Sigma''_{({\rm k})}+
(1-3w)\left(\frac{a'}{a}\right)\Sigma'_{({\rm k})}-
\left\{{\textstyle{3\over2}}\left[1+(2-3w)w\right]
\left(\frac{a'}{a}\right)^2 -{\rm k}^2\right\} \\Sigma_{({\rm
k})}=0\,,  \label{eq:Sig''}
\end{equation}
respectively (see also Appendix).

\subsection{Scale of the induced magnetic field}
The comoving wavenumber of the induced magnetic field depends on
the coherence length of the background field, on the wavelength of
the inducing gravitational radiation and on the interaction angle
between these two sources. To be precise,
\begin{equation}
\ell= {\rm n}\left[1+\left(\frac{\rm k}{\rm n}\right)^2
+2\left(\frac{\rm k}{\rm n}\right)\cos\varphi\right]^{1/2}\,,
\label{eq:ell}
\end{equation}
since ${\rm n}$ takes finite values only. Assuming that ${\rm k}$
and therefore $\ell$ are also finite, the wavelengths
$\lambda_{\tilde{B}}=a/{\rm n}$, $\lambda_{GW}=a/{\rm k}$ and
$\lambda_{B}=a/\ell$ are all well defined and finite. Then,
Eq.~(\ref{eq:ell}) provides the following expression
\begin{equation}
\lambda_B= \lambda_{\tilde{B}}\left[1+
\left(\frac{\lambda_{\tilde{B}}}{\lambda_{GW}}\right)^2
+2\left(\frac{\lambda_{\tilde{B}}}{\lambda_{GW}}\right)
\cos\varphi\right]^{-1/2}\,,  \label{eq:lambdaB}
\end{equation}
for the coherence scale of the induced field. Given that
$0\leq\varphi<\pi/2$, this means
$\lambda_B\leq\lambda_{\tilde{B}}$ always. In particular, we find
$\lambda_B\sim\lambda_{\tilde{B}}$ when
$\lambda_{\tilde{B}}\sim\lambda_{GW}$ and
$\lambda_{\tilde{B}}\ll\lambda_{GW}$, whereas
$\lambda_{GW}\ll\lambda_{\tilde{B}}$ implies
$\lambda_B\ll\lambda_{\tilde{B}}$.

\section{Gravito-magnetic dynamo}
\subsection{Magnetic amplification}
After inflation the only period of low conductivity is that of
early reheating, when the effective equation of state corresponds
to a pressureless fluid. For $p=0$ we have $w=0$,
$a=H_0^2a_0^3\eta^2/4$ and $a'/a=2/\eta$. Then, expressions
(\ref{eq:cB''}) and (\ref{eq:Sig''}) simplify to
\begin{equation}
{\cal B}''_{({\ell})}+ \frac{2}{\eta}{\cal B}'_{({\ell})}+
\ell^2{\cal B}_{({\ell})}=
\frac{8\alpha_1}{\eta^2}\left(\Sigma'_{({\rm k})}
+\frac{1}{\eta}\Sigma_{({\rm k})}\right)  \label{eq:dcB''}
\end{equation}
and
\begin{equation}
\Sigma''_{({\rm k})}+ \frac{2}{\eta}\Sigma'_{({\rm k})}-
\left(\frac{6}{\eta^2}-{\rm k}^2\right)\Sigma_{({\rm k})}=0\,,
\label{eq:dSig''}
\end{equation}
respectively (with $\alpha_1=\kappa^{1/2}\tilde{B}_0^{({\rm
n})}/a_0H_0^2$). Superhorizon-sized gravity waves, with
$\lambda_{GW}\gg\lambda_{H}=1/H$, have ${\rm k}\eta\ll1$ and the
dominant mode in the solution of Eq.~(\ref{eq:dSig''}) is
$\Sigma^{({\rm k})}=\Sigma_0^{({\rm k})}(\eta/\eta_0)^2$.
Substituted into (\ref{eq:dcB''}) the latter gives
\begin{equation}
{\cal B}''_{({\ell})}+ \frac{2}{\eta}{\cal B}'_{({\ell})}+
\ell^2{\cal B}_{({\ell})}= \frac{6\beta_1}{\eta}\,,
\label{eq:dcB''2}
\end{equation}
where $\eta_0^2=4/a_0^2H_0^2$ and $\beta_1=
\kappa^{1/2}a_0\tilde{B}_0^{({\rm n})}\Sigma_0^{({\rm k})}$. This
describes a forced oscillation with a damping effect due to the
expansion. The force depends on the strength of the background
magnetic field and on the gravitationally induced shear at the
beginning of the gravito-magnetic interaction. When $\ell\neq0$ we
obtain
\begin{equation}
{\cal B}^{({\ell})}= {\cal B}^{({\ell})}(\eta)= {\cal
B}_0^{({\ell})}\left[{\rm cos}({\ell}\eta)+{\rm
sin}({\ell}\eta)\right]\left(\frac{\eta_0}{\eta}\right)+
\frac{6\beta_1}{\ell^2\eta}\,,  \label{eq:dcB}
\end{equation}
which on super-Hubble scales (i.e.~for $\ell\eta\ll1$ and ${\rm
cos}({\ell}\eta)+{\rm sin}({\ell}\eta)\simeq1+\ell\eta\simeq1$)
reduces to
\begin{equation}
{\cal B}^{({\ell})}= {\cal B}_0^{({\ell})}
\left(\frac{\eta_0}{\eta}\right)+ \frac{6\beta_1}{\ell^2\eta}\,.
\label{eq:ldcB}
\end{equation}
Finally, recalling that ${\cal
B}^{({\ell})}=\kappa^{1/2}B^{({\ell})}/\Theta$ and using the
relations $\eta^2=4a/H_0^2a_0^3$ and $\Theta=24/H_0^2a_0^3\eta^3$
of the $w=0$ era, Eq.~(\ref{eq:ldcB}) gives\footnote{In~\cite{TDM}
all the solutions were obtained under the assumption that
$\lambda_{GW}\sim\lambda_{\tilde{B}}$. This ensured that
$\lambda_B\sim\lambda_{GW}\sim\lambda_{\tilde{B}}$ (with all three
wavelengths finite - see Eq.~(\ref{eq:ell})) and allowed us to
replace $\lambda_B$ with $\lambda_{\tilde{B}}$ in the final
expressions (e.g.~see Eq.~(21) in~\cite{TDM}). As we will show in
section 6, this special case corresponds to the maximum (resonant)
magnetic amplification. Also, the initial conditions of~\cite{TDM}
assumed $B_0^{(\ell)}=\tilde{B}_0^{({\rm n})}$ instead of setting
$B_0^{(\ell)}$ to zero.}
\begin{equation}
B^{({\ell})}= 9\Sigma_0^{({\rm k})}
\left(\frac{\lambda_B}{\lambda_H}\right)_0^2\tilde{B}_0^{({\rm
n})}\left(\frac{a_0}{a}\right)^2= 9\Sigma_0^{({\rm k})}
\left(\frac{\lambda_B}{\lambda_H}\right)_0^2 \tilde{B}^{({\rm
n})}\,,  \label{eq:ldB2}
\end{equation}
where $\lambda_B$ is the scale of the induced field. Also, since
$B_a$ vanishes in the background (see section 2.2) we have set
$B_0^{(\ell)}=0$. Accordingly, the gravito-magnetic interaction
can lead to a substantial amplification of the B-field when
$9\Sigma_0^{({\rm k})}(\lambda_B/\lambda_H)_0^2\gg1$. For
inflation produced, superhorizon-sized magnetic fields this is a
realistic possibility. In other words, the Maxwell-Weyl coupling
discussed here can act as an effective large-scale dynamo during
the early stages of reheating.

It should be noted that the above results also apply to the
post-recombination universe, provided that the plasma effects are
negligible (e.g.~when ${\rm curl}{\cal J}_a=0$). In that case the
radiation era solution of (\ref{eq:cB''}), (\ref{eq:Sig''}) is
almost identical to Eq.~(\ref{eq:ldB2}) (see Eq.~(25)
in~\cite{TDM}).

\subsection{Gravitationally induced shear}
A common feature in almost all inflationary models is the
production of gravitational radiation with wavelengths extending
over a wide range of scales. In fact, a relic gravity-wave
spectrum is perhaps the only direct signature of inflation that
may still be observable today. The energy density of a linearised
gravity-wave mode produced during a period of de Sitter expansion
is (e.g.~see~\cite{KT})
\begin{equation}
\kappa\rho_{GW}={\textstyle{1\over2}}\int\left[\left(\Delta
h_{+}\right)^2+\left(\Delta h_{\times}\right)^2\right]{\rm
k}^*{\rm d}{\rm k}^*=\frac{2{{\rm
k}^*}^2}{\pi}\left(\frac{H}{m_{Pl}}\right)^2\,.  \label{eq:rhoGW2}
\end{equation}
where ${\rm k}^*$ is the physical wavenumber of the mode. Also,
$\Delta h_{+,\times}=(2/\pi^{1/2})(H/m_{Pl})$ is the mean
fluctuation of the metric perturbation $h_{+,\times}$ and $m_{Pl}$
is the Planck mass~\cite{KT}.  Clearly, $\rho_{GW}\rightarrow0$ as
${\rm k}^*\rightarrow0$.

To proceed further we note that the energy density of
gravitational wave perturbations is related to the magnitude of
the transverse and trace-free part of the shear tensor by
$\kappa\rho_{GW}=\sigma^2$~\cite{MTU}. Then, expression
(\ref{eq:rhoGW2}) takes the form
\begin{equation}
\Sigma=\left(\frac{2}{9\pi}\right)^{1/2}
\left(\frac{\lambda_H}{\lambda_{GW}}\right)
\left(\frac{H}{m_{Pl}}\right)\,,  \label{eq:Sigma}
\end{equation}
where $\Sigma=\sigma/\Theta$ and $\lambda_{GW}=1/{\rm k}^*$. The
above measures the shear anisotropy due to gravitational radiation
of inflationary origin. As expected, the anisotropy depends on the
parameters of the underlying model of inflation (i.e.~on the value
of $H/m_{Pl}$) and it is inversely proportional to the scale of
the wave mode.

\section{Gravito-magnetic resonance}
Hyperhorizon-sized magnetic fields emerge naturally by the end of
inflation, when subhorizon quantum fluctuations in the Maxwell
field are stretched outside the Hubble radius and then freeze-in
as classical electromagnetic waves. At that time the universe is
also permeated by large-scale gravitational waves; the inevitable
prediction of almost all inflationary scenarios. Following
Eq.~(\ref{eq:ldB2}), the effect of the linear interaction between
these two sources depends on the gravitationally induced shear
anisotropy. For inflation produced gravitons the latter is
inversely proportional to their wavelength (see (\ref{eq:Sigma})).
Thus, combining relations (\ref{eq:ldB2}) and (\ref{eq:Sigma}) we
obtain
\begin{equation}
B^{(\ell)}\sim\left(\frac{\lambda_B}{\lambda_H}\right)_0
\left(\frac{\lambda_B}{\lambda_{GW}}\right)_0
\left(\frac{H}{m_{Pl}}\right)_0 \tilde{B}^{({\rm n})}\,.
\label{eq:lB2}
\end{equation}
Note that the zero suffix marks the beginning of the
gravito-magnetic interaction, which here is the end of inflation.
According to expression (\ref{eq:lB2}), we have a substantial
amplification of the geometrically induced B-field if
\begin{equation}
{\cal A}\equiv\left(\frac{\lambda_B}{\lambda_H}\right)_0
\left(\frac{\lambda_B}{\lambda_{GW}}\right)_0
\left(\frac{H}{m_{Pl}}\right)_0\gg1\,,  \label{eq:cA1}
\end{equation}
where ${\cal A}$ may be seen as the amplification factor. Given
that the ratio $(H/m_{Pl})_0$ is fixed by the adopted model of
inflation, the effect of the Maxwell-Weyl coupling depends on the
initial relation between $\lambda_B$, $\lambda_{GW}$ and
$\lambda_H$. In particular, since we are confined to superhorizon
scales, the magnitude of the amplification factor depends
primarily on $\lambda_B$ and $\lambda_{GW}$. These are related to
each other and also to the scale of the background field by
Eq.~(\ref{eq:lambdaB}), which transforms expression (\ref{eq:cA1})
into
\begin{equation}
{\cal A}= {\cal A}(\chi)= 10^{\alpha}
\left(\frac{H}{m_{Pl}}\right)_0
\left[\chi\left(1+\chi^2+2\chi\cos\phi\right)^{-1}\right]\,,
\label{eq:cA2}
\end{equation}
with $\chi=(\lambda_{\tilde{B}}/\lambda_{GW})_0$ by definition.
The latter varies between $0<\chi<\infty$ and determines the
scale-ratio of the two interacting sources. The parameter $\alpha$
determines the coherence length of $\tilde{B}_0$, relative to the
horizon length at the time, according to
$(\lambda_{\tilde{B}}/\lambda_H)_0=10^{\alpha}$. Typically
$\alpha\gg1$, since $(\lambda_{\tilde{B}})_0\gg(\lambda_H)_0$ by
the end of inflation. Once $\alpha$ and $(H/m_{Pl})_0$ are fixed,
the point of maximum amplification is decided by the function
within the brackets. It is then straightforward to show that
${\cal A}(\chi)$ takes its maximum value at $\chi=1$, or
equivalently for $(\lambda_{\tilde{B}})_0=(\lambda_{GW})_0$. In
other words, the maximum magnetic boost is achieved when the two
interacting sources are in resonant coupling. For $\chi=1$ the
amplification factor becomes ${\cal A}={\cal A}_{\rm max}\simeq
10^{\alpha}(H/m_{Pl})_0$. When $\chi\ll1$ or $\chi\gg1$, on the
other hand, expression (\ref{eq:cA2}) ensures that ${\cal
A}\ll{\cal A}_{\rm max}$. Thus, the maximum magnitude of the
gravitationally induced magnetic field is
\begin{equation}
B^{(\ell)}= (B^{(\ell)})_{\rm max}\sim 10^{\alpha}
\left(\frac{H}{m_{Pl}}\right)_0 \tilde{B}^{({\rm n})}\,,
\label{eq:lB3}
\end{equation}
where $\alpha\gg1$. All these mean that the interaction between
inflation produced magnetic seeds and gravitational waves in the
poorly conductive environment of early reheating, can lead to the
resonant amplification of the former. Following (\ref{eq:cA2}),
the maximum magnetic growth occurs at $\chi=1$ irrespective of the
value of $\varphi$. The latter determines only the shape of the
amplification curve. In other words, the gravito-magnetic
resonance is independent of the interaction angle between the two
sources.

Expression (\ref{eq:lB3}) provides the spectrum of the
gravitationally amplified magnetic field at the end of the
resonant-growth phase. The latter occurs during the early stages
of reheating when the electrical conductivity of the cosmic fluid
is low. Once the conductivity has grown beyond a certain
threshold, however, the plasma effects become important (see
section 8). When this happens the electric current term in
Eq.~(\ref{eq:ddotB1}) needs to be accounted for and our analysis
no longer holds. For our purposes, the gravito-magnetic resonance
and the geometric amplification of the induced B-field cease at
that point.

\section{Amplification of inflationary magnetic seeds}
Our results so far have shown that the maximum growth of the
gravitationally induced magnetic field depends on the scale and
the magnitude of the initial seed, as measured at the end of
inflation, and on the adopted inflation model. Given that
inflation has stretched these fields well outside the horizon,
their amplification can be very substantial. In what follows we
will consider three alternative scenarios of magnetogenesis and
calculate the strengths of the resonantly amplified seeds in each
case. Our aim is to obtain a first estimate of the resonant
magnetic growth in each case and to illustrate the potential of
the Maxwell-Weyl coupling as a very efficient early-universe
dynamo.

Large-scale magnetic fields of inflationary origin are generally
extremely weak. Typically, the current energy density of a
primordial field that survived a phase of de Sitter expansion (in
spatially flat FRW universes) is $\rho_B=10^{-104}\lambda_{\rm
Mpc}^{-4}\rho_{\gamma}$, where $\rho_B=B^2/8\pi$, $\rho_{\gamma}$
is the radiation energy density and $\lambda_{\rm Mpc}$ is the
field's physical scale today~\cite{TW}. For a magnetic field with
a coherence length of $\sim10$~kpc, which is the minimum required
for the galactic dynamo to work, the corresponding strength is
roughly $10^{-53}$~G. Such seeds are too weak to support the
dynamo and are therefore treated as astrophysically irrelevant.
However, the interaction of the aforementioned field with gravity
waves during the early stages of reheating can lead to the
resonant amplification of the former according to
Eq.~(\ref{eq:lB3}). Since physical lengths are inversely
proportional to the radiation temperature, a scale of
$\lambda_{\tilde{B}}\sim10$~kpc today translates into
$\lambda_{\tilde{B}}/\lambda_H\sim10^{21}$ at the end of
inflation. The latter is obtained by assuming GUT-scale inflation
with a typical value of $H_0\sim10^{13}$~GeV, which corresponds to
a temperature of approximately $10^{15}$~GeV at the
time.\footnote{At the end of inflation the scale factor
corresponds to a temperature ($T$) given by the formula
$H=(8\pi^{3/2}g_*^{1/2}/\sqrt{90})(T^2/m_{Pl})$, where
$g_*=g_*(T)\simeq10^2$ is the number of the relativistic degrees
of freedom (e.g.~see~\cite{DDPT}).} These mean $\alpha\simeq21$
and $(H/m_{Pl})_0\sim10^{-6}$ for the resonant amplification
parameters of Eq.~(\ref{eq:cA2}). As a result, the associated
maximum-growth factor is of the order of $10^{15}$ and the initial
magnetic seed is amplified to $\sim10^{-38}$~G. Despite this, the
residual field is still some four orders of magnitude below the
minimum required strength of $\sim10^{-34}$~G (see~\cite{DLT}).
Put another way, for workable results we need a stronger initial
seed.

When dealing with spatially flat FRW backgrounds, inflationary
magnetic seeds stronger than $10^{-53}$~G are usually obtained
outside the framework of classical electromagnetic theory. Such an
inflation-based scenario of large-scale magnetogenesis was
recently suggested in~\cite{DDPT}. The proposed mechanism operates
within the standard model, despite breaking the conformal
invariance of the Maxwell field, and produces magnetic seeds of
$\sim10^{-34}$~G on scales of approximately 10~kpc. However,
$10^{-34}$~G is the minimum strength required for the galactic
dynamo to work, and this only in universes dominated by a
dark-energy component. Nevertheless, the interaction of the above
field with gravitational wave perturbations soon after inflation
should boost its amplitude in accordance with Eq.~(\ref{eq:lB3}).
Given the scale of the original seed and using the parameters
of~\cite{DDPT} (i.e.~$H_0\sim10^{13}$~GeV and
$T_0\sim10^{15}$~GeV), the resonant amplification factor is
$10^{15}$ (see also above). The latter brings the residual
magnetic field up to $\sim10^{-19}$~G, which lies very comfortably
within the galactic dynamo requirements for dark energy dominated
cosmologies~\cite{DLT}. Moreover, comoving seeds of $10^{-19}$~G
can sustain the dynamo in conventional universes as well,
especially when the enhancement of the field during the
protogalactic collapse is accounted for.\footnote{In a spherically
symmetric protogalactic collapse, a magnetic field that is
frozen-in with the highly conductive plasma grows as
$B\propto\rho^{2/3}$. This rate, which implies an amplification of
the B-field by three or four orders of magnitude, can increase in
the the more-realistic case of an anisotropically collapsing
protogalaxy due to shearing effects alone~\cite{Z}}

In spatially open universes, standard inflation can produce
magnetic seeds stronger than the typical $10^{-53}$~G fields
without the need to modify Maxwell's theory. In that case the
general relativistic coupling between electromagnetism and the
geometry of the 3-space changes the adiabatic depletion rate of
the magnetic component naturally~\cite{TK,T}. To be precise, on
lengths near the curvature scale, a field that goes through a
period of de Sitter inflation in a perturbed FRW cosmology with
negative spatial curvature decays as $a^{-1}$ instead of $a^{-2}$.
Then, assuming a marginally open universe (i.e.~setting
$1-\Omega\sim10^{-2}$ today), GUT-scale inflation and a reheating
temperature of $\sim10^9$~GeV, one obtains a residual field of
approximately $10^{-35}$~G on a current scale close to $10^4$~Mpc
(see~\cite{TK} for details). For a field on this scale we have
$(\lambda_{\tilde{B}}/\lambda_H)_0\sim10^{27}$, which implies a
resonant amplification factor of the order of $10^{21}$ and a
residual strength of $\sim10^{-14}$~G today. The latter is easily
within the galactic dynamo limits.\footnote{Despite their very
substantial growth the gravitationally amplified magnetic fields
always remain very weak compared to the matter component
(i.e.~$B^2\ll\rho$ at all times). This ensures that our initial
weak-field approximation (see section 2) is never in any doubt and
preserves the symmetries of the FRW background to very high
accuracy.}

The above quoted strengths correspond to resonant amplification.
In other words, we have implicitly assumed that a background
magnetic field of a given length interacts with gravitational
waves of comparable scale. When the two sources have very
different coherence lengths, however, the associated amplification
factors are considerably smaller and the resulting fields much
weaker than those given above (see Eq.~(\ref{eq:cA2})). In
general, of course, the background magnetic seed will interact
with gravity-wave modes of various wavelengths (recall that
$0<\chi<\infty$ in (\ref{eq:cA2})). On these grounds, we expect
the magnitude of the gravitationally induced field to show a
smooth scale-distribution with peak at the point of
gravito-magnetic resonance (i.e.~at $\chi=1$).

\section{Conductivity effects}
\subsection{Low conductivity}
To this point the gravito-magnetic interaction has been free of
current effects, which limits our results to cosmological
environments of very low electrical conductivity. The early
reheating phase of the universe offers such a poorly conductive
stage. As reheating progresses, however, the copious production of
particles continuously increases the conductivity of the universe
and plasma effects become important.

Consider a medium of finite electrical conductivity $\sigma_{\rm
c}$. Phenomenologically, the conductivity effects are accounted
for by means of the electric currents. Using the covariant form of
Ohm's law, in particular, one expresses the 3-current as
(e.g.~see~\cite{J,T})
\begin{equation}
{\cal J}_a= \sigma_{\rm c}E_a\,,  \label{eq:Ohm}
\end{equation}
where $E_a$ is the electric field seen by the observer. Assuming
that the spatial variation of $\sigma_c$ is small, which is a good
approximation on large scales, the above means that ${\rm
curl}{\cal J}_a=\sigma_{\rm c}{\rm curl}E_a$ to linear order and
introduces the conductivity into the right-hand side of
Eq.~(\ref{eq:ddotB1}). Moreover, in a medium of finite
conductivity the magnetic induction equation reads
\begin{equation}
\dot{B}_a+ {\textstyle{2\over3}}\Theta B_a=
\sigma_{ab}\tilde{B}^b- {\rm curl}E_a\,,  \label{eq:fcdotB}
\end{equation}
to first order (e.g.~see~\cite{TB,T}). Note that the time
derivative of the above leads to the linearised wave equation
(\ref{eq:ddotB1}) (see~\cite{T} for details). Employing the
auxiliary relations (\ref{eq:Ohm}) and (\ref{eq:fcdotB}),
Eq.~(\ref{eq:ddotB1}) reads
\begin{equation}
\ddot{B}_a+ {\textstyle{5\over3}}\left(1
+{\textstyle{3\over5}}\frac{{\sigma}_{\rm c}}{\Theta}\right)
\Theta\dot{B}_a+ {\textstyle{1\over3}}\left(1-w
+2\frac{{\sigma}_{\rm c}}{\Theta}\right)\Theta^2B_a- {\rm D}^2B_a=
2\left[\dot{\sigma}_{ab} +{\textstyle{2\over3}}\left(1
+{\textstyle{3\over4}}\frac{{\sigma}_{\rm
c}}{\Theta}\right)\Theta\sigma_{ab}\right]\tilde{B}^b\,,
\label{eq:fcddotB2}
\end{equation}
with the dimensionless ratio $\sigma_{\rm c}/\Theta$ measuring the
electrical conductivity of the expanding background. Hence, the
linear evolution of the gravitationally induced $B$-field depends
on the value of $\sigma_{\rm c}/\Theta$ in a rather involved way.
Nevertheless, the gravito-magnetic interaction proceeds as if the
conductivity were zero as long as $\sigma_{\rm c}/\Theta\ll1$.
This qualitative result was also obtained in~\cite{TW}.

\subsection{High conductivity}
As particle production progresses and the universe heats up, the
conductivity of the cosmic medium increases beyond the
$\sigma_{\rm c}/\Theta\sim1$ threshold and we can no longer ignore
the current term in the right-hand side of Eq.~(\ref{eq:ddotB1}).
Moreover, once the universe enters its standard Big-Bang
evolution, the electrical resistivity is believed to remain very
low~\cite{H2}. When $\sigma_{\rm c}/\Theta\gg1$ the evolution of
the gravitationally induced magnetic field depends largely on the
electrical properties of the fluid (see Eq.~(\ref{eq:fcddotB2})).
The precise role of finite conductivity during reheating lies
beyond the scope of this article, as its study involves highly
sophisticated quantum field theory techniques, it is model
dependent and requires numerical methods to evaluate~\cite{BdVS}.
In what follows we will provide an analytical approach that helps
to outline the implications of a highly conductive cosmological
environment for the proposed gravito-magnetic amplification. For
$w=0$ and $\sigma_c/\Theta\gg1$, which correspond to the late
stages of reheating, Eq.~(\ref{eq:fcddotB2}) gives
\begin{equation}
{\cal B}_{(\ell)}''+ \frac{\sigma_c}{\Theta}\frac{6}{\eta}{\cal
B}_{(\ell)}'+
\left(\frac{\sigma_c}{\Theta}\frac{6}{\eta^2}+\ell^2\right){\cal
B}_{(\ell)}= \frac{8\alpha_1}{\eta^2}\left(\Sigma_{({\rm k})}'
+\frac{\sigma_c}{\Theta}\frac{3}{\eta}\Sigma_{({\rm k})}\right)\,,
\label{eq:hcddotB1}
\end{equation}
where $\alpha_1=\kappa^{1/2}\tilde{B}_0^{({\rm n})}/a_0H_0^2$ and
$\Sigma_{({\rm k})}$ is monitored by (\ref{eq:dSig''}) (see
sections 2.2, 3.1 and also the Appendix). At the
$\sigma_c/\Theta\gg1$ limit, the $\ell$-dependance of the third
term in the left-hand side of the above is only important on
sufficiently small wavelengths (i.e.~for $\ell\eta\gg1$). Here,
however, we are looking at superhorizon scales where ${\rm
n}\eta$, ${\rm k}\eta$ and $\ell\eta\ll1\ll\sigma_c/\Theta$. On
these wavelengths $\Sigma\propto\eta^2$ (see section 3.1) and
expression (\ref{eq:hcddotB1}) reduces to
\begin{equation}
{\cal B}''+ \frac{\sigma_c}{\Theta}\frac{6}{\eta}{\cal B}'+
\frac{\sigma_c}{\Theta}\frac{6}{\eta^2}{\cal B}=
\frac{\sigma_c}{\Theta}\frac{6\beta_1}{\eta}\,,
\label{eq:hcddotB2}
\end{equation}
with $\beta_1=\kappa^{1/2}a_0\Sigma_0\tilde{B}_0$. Note that we
are considering the resonant scenario, with ${\rm k}={\rm
n}=\ell$, which allows us to drop the wavenumber indices in
(\ref{eq:hcddotB2}). Contrary to the case of poor electrical
conductivity (see Eq.~(\ref{eq:dcB''2})), we have arrived at a
scale independent expression. This is due to the highly conductive
environment, which washes out the $\ell$-dependance of
(\ref{eq:hcddotB1}) on sufficiently long wavelengths (i.e.~when
$\ell^2\eta^2\ll\sigma_c/\Theta$). To solve
Eq.~(\ref{eq:hcddotB2}) analytically, consider a brief period of
expansion and assume that in the interval the ratio
$\sigma_c/\Theta$ varies very slowly with time (i.e.~set
$\sigma_c/\Theta\simeq$ constant $\gg1$). Then, since ${\cal
B}=\kappa^{1/2}B/\Theta$, the solution of (\ref{eq:hcddotB2})
approaches the form
\begin{equation}
B=B_0\left(\frac{a_0}{a}\right)^2+
B_0\left(\frac{a_0}{a}\right)^{3\sigma_c/\Theta}+
3\Sigma_0\tilde{B}_0\left(\frac{a}{a_0}\right)\,,
\label{eq:hcB1}
\end{equation}
where $B_0$ can be seen as the gravitationally induced magnetic
field at the onset of the highly conductive regime. The first mode
of the above corresponds to the adiabatic depletion of the field,
while the second carries the plasma effects and decays very
quickly when $\sigma_c/\Theta\gg1$. Hence, for low electrical
resistivity and in the absence of the gravito-magnetic interaction
(i.e.~for $\Sigma_0=0$) we recover the familiar $a^{-2}$-law. The
third mode in (\ref{eq:hcB1}) describes the effect of the
Maxwell-Weyl coupling on the $B$-field. Compared to the low
conductivity case (e.g.~see results (\ref{eq:ldcB}) or
(\ref{eq:lB2})), there is no scale dependence and the
gravito-magnetic resonance has been completely suppressed.
Therefore, as the reheating of the universe progresses and the
electrical resistivity of the cosmic medium drops, the large-scale
effects of the Maxwell-Weyl resonance should fade away. This
suggests that the proposed gravito-magnetic dynamo is only
effective at the early stages of reheating. Similar results,
showing an analogous damping of electromagnetic modes in highly
conductive environments, have been obtained in the past. More
specifically, numerical calculations of the magnetic amplification
due to the parametric resonance of the field with an oscillating
complex scalar field during preheating, showed a very substantial
decrease in the magnetic growth with increasing electrical
conductivity~\cite{BPTV}.

Although the plasma effects may have overwhelmed the
gravito-magnetic resonance, the third mode of Eq.~(\ref{eq:hcB1})
also shows that the Maxwell-Weyl coupling slows down the decay
rate of the field. Interestingly, the same effect was also
obtained through the relativistic coupling of the $B$-field with
the spatial curvature of an open FRW cosmology~\cite{TK}. This
time, however, the geometrically induced superadiabatic magnetic
amplification is not efficient. Indeed, ignoring the fast decaying
second mode in (\ref{eq:hcB1}), the latter reads
\begin{equation}
B=\left[B_0+3\Sigma_0\tilde{B}_0\left(\frac{a}{a_0}\right)\right]
\left(\frac{a_0}{a}\right)^2\,.  \label{eq:hcB2}
\end{equation}
Amplification is therefore achieved only when
$3\Sigma_0\tilde{B}_0(a/a_0)>B_0$. Typically, $\Sigma_0\ll10^{-6}$
(see section 3.2) and $\tilde{B}_0\leq B_0$, which implies that
the above given condition is satisfied at late times only
(i.e.~for $a/a_0\gg1$). As the time interval of the interaction
increases, however, the assumption that $\sigma_c/\Theta\simeq$
constant becomes more difficult to maintain and this affects the
accuracy of our results. Having said that, the same
superadiabatic-type of magnetic amplification is also observed at
the infinite conductivity limit (see footnote 3 in~\cite{TDM}).
All these raise the intriguing possibility of a change in the
adiabatic $a^{-2}$-law due to gravity-wave effects alone and
irrespective of the conductivity of the cosmological environment.

\section{Discussion}
Inflation can naturally achieve superhorizon correlations from
small-scale microphysics. This property has been exploited by
several authors in order to produce primordial magnetic fields
with the desired large coherence lengths. The drawback of
inflation is the dramatic dilution of the magnetic energy density
during the accelerated expansion phase. For a field that survived
inflation and spans a scale of $\sim10$~kpc today, the typical
strength is roughly $10^{-53}$~G. On that scale the minimum
required strength for the galactic dynamo to work is $10^{-34}$~G,
assuming that the universe is dark-energy dominated. Otherwise the
magnetic seed should be at least as strong as $\sim10^{-23}$~G.
The most common solution to the strength problem is by slowing
down the adiabatic, $a^{-2}$ decay rate of the B-field. When
dealing with spatially flat FRW backgrounds, this usually means
breaking the conformal invariance of the Maxwell field and in the
majority of cases it is achieved outside the standard model.

Inflation also produces a background of large-scale gravitational
radiation. The interaction of these waves with inflationary
produced magnetic seeds soon after the end of inflation was first
considered in~\cite{TDM}. The initial results argued for a very
significant growth, by many orders of magnitude, of the primordial
field. Here, we have revisited this gravito-magnetic interaction
in an attempt to understand and explain the physics of the
amplification mechanism further. Our analysis has revealed that
the very strong magnetic growth found in~\cite{TDM}, reflects the
resonant coupling of the two interacting sources in cosmological
environments of poor electrical conductivity. We have shown, in
particular, that the maximum amplification the B-field occurs
always when the coherence scale of the latter coincides with the
wavelength of the inducing gravitational radiation.

The proposed Maxwell-Weyl interaction and the resulting
amplification mechanism are rather simple in concept. At the end
of inflation the universe is permeated by large-scale gravity
waves and by a very weak, large-scale primordial magnetic field.
The general relativistic interaction of these two sources during
early reheating leads to a gravitationally induced magnetic
component. When the associated scales are comparable this field is
resonantly amplified. In general, of course, the original magnetic
seed will interact with gravitational radiation of various
wavelengths. This means that the strength of the induced field
will have a smooth scale-dependent spectrum with peak at the point
of resonance. The maximum strength of the geometrically amplified
magnetic component is determined at the onset of the
gravito-magnetic interaction. Once the parameters of the adopted
inflationary model are fixed, the resonant growth factor is
proportional to the scale of the initial field. This makes the
proposed amplification mechanism particularly effective when
operating on superhorizon-sized magnetic seeds. In particular, for
a field with current physical scale close to 10~kpc, which is the
minimum required for the dynamo to work, the resonant growth is of
the order of $10^{15}$. Although very substantial, such a boost
cannot bring the typical inflation-produced magnetic field of
$\sim10^{-53}$~G (see~\cite{TW}) within the galactic dynamo
requirements. Nevertheless, when applied to seeds of
$\sim10^{-34}$~G and $\sim10^{-35}$~G, like those produced
in~\cite{DDPT} and~\cite{TK} for example, the gravito-magnetic
resonance leads to residual fields of $\sim10^{-19}$~G and
$\sim10^{-14}$~G respectively. The latter can support the galactic
dynamo even in conventional universes with zero dark-energy
contribution.

The resonant amplification of the initial seed field by many
orders of magnitude, makes the proposed gravito-magnetic coupling
a very promising early-universe dynamo. Given that, it is worth
looking into the specifics of the mechanism in more detail. A key
issue is the role of electrical conductivity near and beyond the
$\sigma_{\rm c}/\Theta\sim1$ threshold. Here we found that at the
$\sigma_c/\Theta\simeq$ constant $\gg1$ limit the Maxwell-Weyl
resonance is suppressed, which is in qualitative agreement with
analogous earlier studies (e.g.~see~\cite{M,BPTV}). One could
improve on this result by adopting a specific model for the
conductivity of the reheating era. The nature of adopted
inflationary scenario and of the associated reheating process is
also an issue. Non-oscillatory models, for example, may provide a
longer period of low electrical conductivity and an enhanced
gravity-wave spectrum. Another key question is the magnetic
backreaction on the gravity waves themselves and its potential
impact on the amplification mechanism itself. This issue acquires
particular interest in view of the work of~\cite{CD}. There
large-scale stochastic magnetic fields were found capable of
efficiently converting their energy into gravitational radiation,
as they re-enter the cosmological horizon. It is conceivable that
the simultaneous study of the two processes will point towards a
preferred saturation level for the combined gravito-magnetic
interaction.

\section*{Acknowledgements}
The author wishes to thank Axel Brandenburg, Kostas Dimopoulos,
Hidekazu Hanayama, Alejandra Kandus, Roy Maartens and Shinji
Tsujikawa for helpful discussions and comments.

\section*{Appendix}
\subsection*{Expansion-normalised gravito-magnetic equations}
The linear evolution of the magnetic mode $B_{(\ell)}$ induced by
gravity wave perturbations on a weakly magnetised, spatially flat
FRW universe is governed by the system (see Eqs.~(\ref{eq:ddotB2})
and (\ref{eq:ddotsig2}) in section 2.2)
\begin{equation}
\ddot{B}_{({\ell})}+
{\textstyle{5\over3}}\Theta\dot{B}_{({\ell})}+
\left[{\textstyle{1\over3}}(1-w)\Theta^2
+\frac{\ell^2}{a^2}\right]B_{({\ell})}= 2\left(\dot{\sigma}_{({\rm
k})}+ {\textstyle{2\over3}}\Theta\sigma_{({\rm
k})}\right)\tilde{B}_0^{({\rm n})}\left(\frac{a_0}{a}\right)^2\,,
\label{eq:apddotB}
\end{equation}
\begin{equation}
\ddot{\sigma}_{({\rm k})}+
{\textstyle{5\over3}}\Theta\dot{\sigma}_{({\rm k})}+
\left[{\textstyle{1\over6}}(1-3w)\Theta^2+\frac{{\rm
k}^2}{a^2}\right]{\sigma}_{({\rm k})}=0\,.  \label{eq:apddotsigma}
\end{equation}
Here $\tilde{B}_0^{({\rm n})}$ is the background field,
$\sigma_{({\rm k})}$ is the shear anisotropy due to the
gravitational waves and the zero suffix indicates the beginning of
the gravito-magnetic interaction. To zero perturbative order the
background expansion is monitored by the Friedmann and the
Raychaudhuri equations, given by expressions (\ref{eq:Fried-Ray}a)
and (\ref{eq:Fried-Ray}b) respectively. When combined these reduce
Raychaudhuri's formula to
\begin{equation}
\dot{\Theta}=-{\textstyle{1\over2}}(1+w)\Theta^2\,.
\label{eq:apRay}
\end{equation}

Consider the expansion-normalised, dimensionless variables ${\cal
B}_{(\ell)}=\kappa^{1/2}B_{(\ell)}/\Theta$ and $\Sigma_{({\rm
k})}=\sigma_{({\rm k})}/\Theta$ defined in section 2.2. Using
(\ref{eq:apRay}) we obtain the auxiliary relations
\begin{eqnarray}
\kappa^{1/2}\dot{B}_{(\ell)}&=& \Theta\dot{\cal B}_{(\ell)}-
{\textstyle{1\over2}}(1+w)\Theta^2{\cal B}_{(\ell)}\,,
\label{eq:apdotB1}\\ \kappa^{1/2}\ddot{B}_{(\ell)}&=&
\Theta\ddot{\cal B}_{(\ell)}- (1+w)\Theta^2\dot{\cal B}_{(\ell)}+
{\textstyle{1\over2}}(1+w)^2\Theta^3{\cal B}_{(\ell)}\,,
\label{eq:apddotB1}
\end{eqnarray}
between the proper-time derivatives of ${\cal B}_{(\ell)}$ and
$B_{(\ell)}$. Similarly, the first and second derivatives of
$\sigma_{({\rm k})}$ give
\begin{eqnarray}
\dot{\sigma}_{({\rm k})}&=& \Theta\dot{\Sigma}_{({\rm k})}-
{\textstyle{1\over2}}(1+w)\Theta^2{\Sigma}_{({\rm k})}\,,
\label{eq:apdotsigma1}\\ \ddot{\sigma}_{({\rm k})}&=&
\Theta\ddot{\Sigma}_{({\rm k})}- (1+w)\Theta^2\dot{\Sigma}_{({\rm
k})}+ {\textstyle{1\over2}}(1+w)^2\Theta^3{\Sigma}_{({\rm k})}\,.
\label{eq:apddotsigma1}
\end{eqnarray}
Results (\ref{eq:apdotB1})-(\ref{eq:apddotsigma1}) transform
expressions (\ref{eq:apddotB}) and (\ref{eq:apddotsigma}) into
\begin{equation}
\ddot{\cal B}_{(\ell)}+ {\textstyle{1\over3}}(2-3w)\Theta\dot{\cal
B}_{(\ell)}- \left[{\textstyle{1\over6}}(1-3w)w\Theta^2
-\frac{\ell^2}{a^2}\right]{\cal B}_{(\ell)}=
2\kappa^{1/2}\left[\dot{\Sigma}_{({\rm k})}
+{\textstyle{1\over6}}(1-3w)\Theta\Sigma_{({\rm
k})}\right]\tilde{B}_0^{({\rm n})}\left(\frac{a_0}{a}\right)^2\,,
\label{eq:apddotcB1}
\end{equation}
and
\begin{equation}
\ddot{\Sigma}_{({\rm k})}+
{\textstyle{1\over3}}(2-3w)\Theta\dot{\Sigma}_{({\rm k})}-
\left\{{\textstyle{1\over6}}[1+(2-3w)w]\Theta^2 -\frac{{\rm
k}^2}{a^2}\right\}{\Sigma}_{({\rm k})}=0\,,
\label{eq:apddotSigma1}
\end{equation}
respectively.

The final step is to introduce the conformal time variable $\eta$,
with $\dot{\eta}=1/a$ by definition. Then $\Theta=3a'/a^2$, where
the prime indicates conformal-time derivatives. Accordingly,
\begin{equation}
\dot{\cal B}_{(\ell)}=\frac{1}{a}{\cal B}'_{(\ell)} \hspace{15mm}
{\rm and} \hspace{15mm} \ddot{\cal B}_{(\ell)}=\frac{1}{a^2}{\cal
B}''_{(\ell)}- \frac{a'}{a^3}{\cal B}'_{(\ell)}\,,
\label{eq:apcBder}
\end{equation}
with exactly analogous expressions for $\dot{\Sigma}_{({\rm k})}$
and $\ddot{\Sigma}_{({\rm k})}$ respectively. These relations
recast Eqs.~(\ref{eq:apddotcB1}), (\ref{eq:apddotSigma1}) in terms
of conformal-time derivatives as
\begin{eqnarray}
{\cal B}''_{({\ell})} + (1-3w)\left(\frac{a'}{a}\right){\cal
B}'_{({\ell})} -
\left[{\textstyle{3\over2}}(1-3w)w\left(\frac{a'}{a}\right)^2
-\ell^2\right]{\cal B}_{({\ell})}&=& \nonumber\\
2\kappa^{1/2}a\left[\Sigma'_{({\rm k})}
+{\textstyle{1\over2}}(1-3w)\left(\frac{a'}{a}\right)\Sigma_{({\rm
k})}\right]\tilde{B}_0^{(n)}\left(\frac{a_0}{a}\right)^2
\label{eq:apddotcB2}
\end{eqnarray}
and
\begin{equation}
\Sigma''_{({\rm k})}+
(1-3w)\left(\frac{a'}{a}\right)\Sigma'_{({\rm k})}-
\left\{{\textstyle{3\over2}}\left[1+(2-3w)w\right]
\left(\frac{a'}{a}\right)^2 -{\rm k}^2\right\}\Sigma_{({\rm
k})}=0\,,  \label{eq:apddotSigma2}
\end{equation}
respectively (compare to formulae (\ref{eq:cB''}),
(\ref{eq:Sig''}) of section 2.2).

\end{document}